\documentclass[aps,prl,reprint,nofootinbib,superscriptaddress]{revtex4-1}
\usepackage{amssymb}

\usepackage{tabularx}
\usepackage{subfigure}
\usepackage{amsmath,amssymb}
\usepackage{latexsym}
\usepackage{psfrag}
\usepackage{epsfig}
\usepackage{graphicx,subfigure}
\usepackage{array}
\usepackage{tabularx}
\usepackage{hyperref}

\usepackage{multirow}
\usepackage{epstopdf}
\usepackage{graphicx}

\usepackage{algorithm}

\begin{document}
\title{Effective quantum state reconstruction using compressive sensing in NMR quantum computing}
\author{J. Yang}
\affiliation{Department of Automation,University of Science and Technology of China, Hefei 230027, P. R. China}%
\author{S. Cong}
\email{scong@ustc.edu.cn}
\affiliation{Department of Automation,University of Science and Technology of China, Hefei 230027, P. R. China}%
\author{X. Liu}
\affiliation{Department of Physics, University of Science and Technology of China, Hefei 230026, P. R.China}%
\author{Z. Li}
\affiliation{Department of Physics, University of Science and Technology of China, Hefei 230026, P. R.China}%
\author{K. Li}
\affiliation{Imperial College London, MRC Institute of Medical Sciences, London, W12 0NN, UK}

\begin{abstract}
The number of measurements required to reconstruct the states of quantum systems increases exponentially with the quantum system dimensions, which makes the state reconstruction of high-qubit quantum systems have a great challenge in physical quantum computing experiments. Compressive sensing (CS) has been verified as a effective technique in the reconstruction of quantum state, however, it is still unknown that if CS can reconstruct quantum states given the less data measured by nuclear magnetic resonance (NMR). In this paper, we propose an effective NMR quantum state reconstruction method based on CS. Different from the conventional CS-based quantum state reconstruction, our method uses the actual observation data from NMR experiments rather than the data measured by the Pauli operators. We implement measurements on quantum states in practical NMR computing experiments and reconstruct states of 2,3,4 qubits using fewer number of measurements, respectively. The proposed method is easy to implement and performs more efficiently with the increase of the system dimension size. The performance reveals both efficiency and accuracy, which provides an alternative for the quantum state reconstruction in practical NMR.
\end{abstract}

\maketitle

\section{I. INTRODUCTION}

Nuclear magnetic resonance (NMR) is one of the most promising physical methods to realize quantum computing, which has attracted tremendous interest in both the physics and information science community \citep{lab1,lab2,lab3,lab4}. In practical NMR quantum computing experiments, the reconstruction of quantum states occupies an important position. Conventional quantum state tomography (QST) is a common method for NMR quantum state reconstruction (QSR)\citep{lab5,lab6,lab7}, which requires complete measurements of the quantum state to be reconstructed. For an $n$-qubit state $\rho$, the number of complete measurements is ${d^2} = {4^n}$. This number increases exponentially with $n$, and makes the reconstruction work of the high-qubit NMR state becomes extremely difficult as $n$ is large. In order to reduce the number of measurements, people sometimes use local quantum tomography \cite{lab8} to reconstruct the states. However, local quantum tomography requires a sufficient amount of prior information before reconstructing, which does not have universal applicability. Therefore, the reconstruction of high-qubit NMR quantum states has great challenge.

Compressed sensing (CS) \cite{lab9} has attracted a great interest as an effective approach of recovering sparse signals. This approach is now widely applied in many fields, such as image processing \cite{lab10,lab12}, wireless communication \cite{lab11}, nuclear magnetic resonance imaging (MRI) \citep{lab13,lab14} and NMR spectroscopy \citep{lab15,lab16}. CS also provides a new idea to solve the problem of high-qubit quantum state reconstruction. People have performed accurate quantum state reconstruction in physical systems such as photon \citep{lab17,lab18,lab19}  and ion trap \cite{lab20}. However, it is still not clear that if CS can reconstruct quantum states provided measurement data from NMR, because in NMR the measurements of QSR are obtained in a different way. In practical NMR computing experiments, the signal of each measurement is sampled in the time-domain and then transferred into a frequency-domain spectrum. The spectrum contains a number of resonance peaks, and each peak is associated with an observable ${O_i}$ of the system. The observables of the same spectrum constitute an NMR observable group, which can be measured simultaneously. When reconstructing an actual NMR state $\rho$ based on CS, due to the particularity of practical NMR measurements, the observables should be sampled in units of observable groups rather than individual observables ${O_i}$ , which is different from the sampling method in conventional CS-based QST \citep{lab21,lab22,lab23}. Moreover, the CS optimization problem requires an appropriate optimization algorithm. The most commonly used optimization algorithms in CS are LS \cite{lab24}, Dantzig \cite{lab25}, gradient projection \cite{lab26} and so on. Li and Cong first applied the ADMM algorithm to QSR and showed that the ADMM algorithm has better performance than the commonly used algorithms \citep{lab27,lab28}. FP-ADMM algorithm was proposed by Zheng et al. \cite{lab29}, which combines the fixed point idea and ADMM algorithm and improves further the efficiency of the reconstruction.

In this paper, we show that CS is an effective technique for the NMR quantum state reconstruction. To our knowledge, we carry out the first experimental reconstruction of NMR states by using CS. We theoretically prove that CS can be applied to the reconstruction of actual NMR quantum states, and give the detailed reconstruction steps, which combines CS and the characteristics of the practical NMR measurement. FP-ADMM is used as the optimization algorithm to solve the CS optimization problem. We experimentally verify the proposed method by the reconstruction of actual NMR states with 2,3,4 qubits and analyze the effects of factors on the reconstruction performance. The experimental results show that the proposed method can effectively reconstruct the NMR quantum state using only a small amount of measurement data. The reconstruction performance reveals both efficiency and accuracy with the increase of system dimension size. The proposed method is easily applicable to higher qubits for any NMR low-rank quantum states.

The structure of this paper is as follows. In Sec. II, after a brief introduction to the practical NMR measurement method, the CS theory and the FP-ADMM algorithm, we prove our reconstruction method and give specific steps of the method. In Sec. III, the experimental reconstruction results are shown. We perform reconstruction of actual 2,3,4 qubits NMR states, respectively, and analyze the reconstruction performance through contrast experiments. The conclusion of this paper is given in Sec. IV.

\section{II. QUANTUM STATE RECONSTRUCTION BASED ON COMPRESSIVE SENSING AND ACTUAL NMR OBSERVATION DATA}

As an indirect measurement process, the system to be measured in NMR is an $n$-qubit quantum state consisting of spin nuclei in the sample solution under a constant $z$-direction magnetic field ${B_0}$. There is a magnetic moment of the spin nuclei in the magnetic field, whose direction is same as ${B_0}$ and magnitude is proportional to the angular momentum of the spin. The external control field is a radio frequency (RF) pulse magnetic field on the $x-y$ plane. When applying an RF pulse consisting of a plurality of resonant frequencies to the sample solution, the nuclei absorbs the energy of the RF pulse, and the angle between the magnetic moment and ${B_0}$ changes, leading to a Larmor precession of the nuclei. There is an induction coil winding on the surface of the sample solution, and the nuclear precession results in a free induction decay current signal $s(t)$ in the induction coil: $s(t) = \sum\nolimits_i {{{\bf{M}}_0}{{\mathop{\rm e}\nolimits} ^{i{\Omega _i}t}}{e^{ - t/{T_2}}}} $, where $t$ stands for time, ${\Omega _i}$ denotes the resonant frequencies, and $i$ is the flag, ${{\bf{M}}_0}$ is the value of fixed RF field magnetization intensity vector, and ${T_2}$ represents the transverse relaxation time. $S(\omega )$ is a frequency-domain spectrum which is obtained from the Fourier transform of $s(t)$: $S(\omega ) = \int_0^\infty  {s(t){e^{ - i\omega t}}dt}  = A(\omega ) + iB(\omega )$ £¬where $\omega$ stands for the frequency, and $A(\omega ) = \sum\nolimits_i {{\bf{K}}{{{T_2}^{ - 1}} \mathord{\left/
 {\vphantom {{{T_2}^{ - 1}} {{{(\omega  - {\Omega _i})}^2} + {T_2}^{ - 2}}}} \right.
 \kern-\nulldelimiterspace} {{{(\omega  - {\Omega _i})}^2} + {T_2}^{ - 2}}}} $ and $B(\omega ) = \sum\nolimits_i {{\bf{K}}{{(\omega  - {\Omega _i})} \mathord{\left/
 {\vphantom {{(\omega  - {\Omega _i})} {{{(\omega  - {\Omega _i})}^2} + {T_2}^{ - 2}}}} \right.
 \kern-\nulldelimiterspace} {{{(\omega  - {\Omega _i})}^2} + {T_2}^{ - 2}}}} $ are the real and imaginary parts of $S(\omega )$ , respectively. The spectrum of $A(\omega )$ and $B(\omega )$ near the resonant frequency ${\Omega _i}$  are resonance peaks, with the peaks of $A(\omega )$ being absorption peaks and that of $B(\omega )$ being symmetric dispersion peaks.

When measuring an $n$-qubit quantum state $\rho$ whose dimension is $d = {2^n}$, each resonance peak in $S(\omega )$ corresponds to an observable ${O_i}$, and the observation value of ${O_i}$ is proportional to the area of the signals in the corresponding peak of $A(\omega )$:
\begin{equation}\label{eq1}
\left\langle {{O_i}} \right\rangle  = \frac{1}{{{{\bf{P}}_{\bf{0}}}}}\int_{{\Omega _i} - \Delta \omega }^{{\Omega _i} + \Delta \omega } {A(\omega )d\omega },
\end{equation}
where ${{\bf{P}}_{\bf{0}}}$ is the scaling factor which can be determined by the peak$'$s area of the eigenstate in the same sample solution, and $\Delta \omega $  is a fixed range value, which ensures all signals of the selected formant are included in the frequency range $\left[ {{\Omega _i} - \Delta \omega ,{\Omega _i} + \Delta \omega } \right]$.

As the spectrum $S(\omega )$ contains $d$ resonance peaks, the observation data of the corresponding $d$ observables are obtained simultaneously in one NMR measurement. Such $d$ observables constitute an NMR observable group, defined as $\left\{ {O_j^k} \right\} = \left\{ {O_1^k,O_2^k,...,O_d^k} \right\}$, where $j = 1,2,...,d$, and $k = 1,2,...,v$ is the serial number of the group. Here $v$ denotes the total number of the observable groups which is determined by the composition of the experimental sample and the actual measurement scheme. For example, $O_j^k$ represents the $j$-th observable in the $k$-th group and is also expressed in the subscript form as $O_j^k=O_{kj}$. $\left\{ {{O_i}} \right\}$ (${i = 1,2,...,vd}$) is the set of all the observables. In practical NMR experiments, people design a measurement scheme of $v$ different NMR observable groups to measure the complete observables of $\rho$, with some inevitably repetitive or linearly related observables in different groups, meaning that the total observables of $\left\{ {{O_i}} \right\}$ are over-complete for $\rho$.

Because the observables of $\left\{ {{O_i}} \right\}$ are over-complete, a conventional method for NMR QST is to perform the following transformation: based on the $d$ observables $O_j^k$ in each group $\left\{ {O_j^k} \right\}$, $\left\{ {O_j^k} \right\}$ can be transformed into a set of measurement operators $\left\{ {M_j^k} \right\} = \left\{ {M_1^k,M_2^k,...,M_d^k} \right\}$  by $M_j^k = \sum\nolimits_i^{{2^{n - 1}}} {{h_{ij}}O_j^k} $, where $M_j^k$ is an $n$-qubit Pauli operator that is the tensor product of Pauli matrices $\{I, X, Y, Z\} = \left\{ \left( \begin{array}{cc}1 & 0 \\ 0 & 1 \end{array} \right), \left(\begin{array}{cc}0& 1 \\ 1 & 0 \end{array}  \right), \left(\begin{array}{cc}0 & -i \\ i & 0 \end{array} \right) ,\left(\begin{array}{cc}1 & 0 \\ 0 & -1 \end{array}   \right)\right\}$, and ${h_{ij}}$ are the elements in the transformation matrix ${\bf{H}}$ of $\left\{ {O_j^k} \right\}$ and $\left\{ {M_j^k} \right\}$. Each column of ${\bf{H}}$ represents a linear transformation from $\left\{ {O_j^k} \right\}$ to one operator of $\left\{ {M_j^k} \right\}$. The observation values $\left\{ {\left\langle {O_j^k} \right\rangle } \right\}$ can also be transformed to measured values $\left\{ {\left\langle {M_j^k} \right\rangle } \right\}$. It is easy for $\left\{ {M_j^k} \right\}$ to remove all the repetitions and get a complete set of measurement operators $\left\{ {{M_m}} \right\}$ ($m = 1,2,...,{d^2}$), thus one can directly calculate the reconstructed density matrix $\hat \rho $  of the state  $\rho $  with $\left\{ {{M_m}} \right\}$ and $\left\{ {\left\langle {{M_m}} \right\rangle } \right\}$: $\hat \rho  = \frac{1}{{{2^n}}}\sum\nolimits_{m = 1}^{{4^n}} {\left( {\left\langle {{M_m}} \right\rangle  \cdot {M_m}} \right)} $. However, as the number of qubits increases, the number of measurement operators required for quantum tomography increases exponentially, and the corresponding actual NMR measurement becomes extremely cumbersome.

Here we propose an effective reconstruction method of the actual NMR quantum states based on CS, in which we directly use observables $\left\{ {{O_i}} \right\}$ but not the transformed measurement operators $\left\{ {{M_m}} \right\}$. One necessary condition of using CS in QSR is that the density matrix of $\rho$ should be low-rank. That is, the rank of $\rho$ is much less than its dimension: $r \ll d$. In practical NMR quantum computing experiments, the quantum state to be reconstructed is mostly pure or nearly pure, which satisfies the low-rank condition. Therefore, CS can be applied to the reconstruction of actual NMR quantum states. The reconstruction process can be described as to solve the following convex optimization problem:
\begin{equation}\label{eq2}
\min {\left\| \rho  \right\|_*}, \ \ \text{s.t.} \ {\bf{y}} = {\bf{A}} \cdot {\mathop{\rm vec}\nolimits} (\rho ),
\end{equation}
where ${\left\| \rho  \right\|_*}$ is the nuclear norm of $\rho$, which equals to the sum of singular values, ${\mathop{\rm vec}\nolimits} ( \cdot )$  represents the transformation from a matrix to a vector by stacking the matrix$'$s columns in order on the top of one another. The sampling matrix ${\bf{A}}$ is the matrix form of the all the sampled observables ${O_i}$, and the sampling vector ${\bf{y}}$ is the vector form of the corresponding observation values $\left\langle {{O_i}} \right\rangle $.

Considering the measurement method in NMR experiment, the observable groups $\left\{ {O_j^k} \right\}$  and the corresponding actual observation values $\left\{ {\left\langle {O_j^k} \right\rangle } \right\}$  are randomly sampled for the CS-QSR. Because the observation values are sampled in groups, here we defined a new sampling rate as
\begin{equation}\label{eq3}
{\eta _g} = {g \mathord{\left/
 {\vphantom {g v}} \right.
 \kern-\nulldelimiterspace} v},
\end{equation}
where $g$ is the number of the sampled groups, and $v$ is the total number of groups. It is worth mentioning that ${\eta _g}$  is different from the general sampling rate ${\eta _m} = {m \mathord{\left/
 {\vphantom {m {{d^2}}}} \right.
 \kern-\nulldelimiterspace} {{d^2}}}$ , where $m$ and ${d^2}$ represent the sampled number and total number of the measurement operators $\left\{ {{M_m}} \right\}$, respectively, and ${\eta _g},{\eta _m} \in [0,1]$.

Without loss of generality, assuming that the randomly sampled serial number is from $1$ to $g$, then ${\bf{A}}$ and ${\bf{y}}$ in NMR can be written as
\begin{equation}\label{eq4}
{\bf{A}} = \left( \begin{array}{l}
{\mathop{\rm vec}\nolimits} {(\left\{ {O_j^1} \right\})^T}\\
{\mathop{\rm vec}\nolimits} {(\left\{ {O_j^2} \right\})^T}\\
\mathop {}\nolimits_{}^{}  \vdots \\
{\mathop{\rm vec}\nolimits} {(\left\{ {O_j^g} \right\})^T}
\end{array} \right)/\sqrt d,
\end{equation}
and
\begin{equation}\label{eq5}
{\bf{y}} = {\left( {\left\{ {\left\langle {O_j^1} \right\rangle } \right\},\left\{ {\left\langle {O_j^2} \right\rangle } \right\}, \cdot  \cdot  \cdot ,\left\{ {\left\langle {O_j^g} \right\rangle } \right\}} \right)^T},
\end{equation}
where ${\mathop{\rm vec}\nolimits} {(\left\{ {O_j^k} \right\})^T}$ represents the transformation from the $d$ observables of $\left\{ {O_j^k} \right\}$ to $d$ horizontal vectors arranged in vertical order: \begin{footnotesize}${\mathop{\rm vec}\nolimits} {(\left\{ {O_j^k} \right\})^T} = \left( \begin{array}{l}
{\mathop{\rm vec}\nolimits} {(O_1^k)^T}\\
{\mathop{\rm vec}\nolimits} {(O_2^k)^T}\\
\mathop {}\nolimits_{}^{}  \vdots \\
{\mathop{\rm vec}\nolimits} {(O_d^k)^T}
\end{array} \right)$\end{footnotesize}, and ${\bf{y}}$  is the vector of the observation values corresponding to the observables of ${\bf{A}}$. In this case, the optimization problem (2) is an equation group composed of $g \times d$  equations. It should be noted that, since the total observables are over-complete, there may be some repeating equations in (2), but this repetition does not affect the solution of (2).

Candes et al. proved that, if the sampling matrix ${\bf{A}}$ satisfies the rank restricted isometry property (RIP) \cite{lab30}, the convex optimization problem (2) has a unique optimal solution equaling to the true density matrix \cite{lab31}. It is proved that the sampling matrix ${\bf{A}}$ consisting of randomly sampled Pauli measurement operators satisfies rank RIP with very high probability \cite{lab25}. Since the transformation between the operators of $\left\{ {O_j^k} \right\}$ and $\left\{ {M_j^k} \right\}$ is linear, if the sampling matrix ${{\bf{A}}_M}$ consisting of $g$ different Pauli measurement operator groups $\left\{ {M_j^k} \right\}$ satisfies rank RIP, then the sampling matrix ${{\bf{A}}_O}$ that consists of corresponding $g$ observable groups $\left\{ {O_j^k} \right\}$ also satisfies rank RIP. This means, in theory, our method sampling the observable groups $\left\{ {O_j^k} \right\}$ is applicable to the reconstruction of actual NMR quantum state $\rho$.

In this paper, we use the FP-ADMM algorithm proposed by Zheng et al.\cite{lab29} to solve the optimization problem (2). The iterative steps of the FP-ADMM algorithm are as follows:
\begin{footnotesize}
\begin{equation}\label{eq6}
\begin{cases}
\rho _1^{k\!+\!1{\kern 1pt} {\kern 1pt} }{\kern 1pt}\!\!=\!\!{D_{\delta {\textstyle{1 \over \mu }}}}({\mathop{\rm mat}\nolimits} ((I\!-\!\delta {{\bf{A}}^\dag }{\bf{A}}){\mathop{\rm vec}\nolimits} (\rho _1^{k{\kern 1pt} })\!+\!\delta {{\bf{A}}^\dag }({\bf{y}}\!-\!{\bf{A}} \cdot {\mathop{\rm vec}\nolimits} ({S^k})\!-\!\frac{{{Y^k}}}{\mu })))\\
\rho _{}^{k + 1} = \frac{1}{2}(\rho _1^{k + 1} + {{\left( {\rho _1^{k + 1}} \right)}^\dag }){\rm{                                                                  }}\\
S{\kern 1pt} _{}^{k\!+\!1}{\kern 1pt} {\kern 1pt}\!\!=\!\!{S_{\delta {\textstyle{\lambda  \over \mu }}}}({\mathop{\rm mat}\nolimits} ((I\!-\! \delta {{\bf{A}}^\dag }{\bf{A}}){\mathop{\rm vec}\nolimits} ({S^k})\!+\! \delta {{\bf{A}}^\dag }({\bf{y}}\!-\!{\bf{A}} \cdot {\mathop{\rm vec}\nolimits} ({\rho ^{k + 1}})\!-\!\frac{{{Y^k}}}{\mu })))\\
{Y^{k + 1}} = {Y^k} + \mu \left[ {{\bf{A}} \cdot {\mathop{\rm vec}\nolimits} ({\rho ^{k + 1}} + {S^{k + 1}}) - {\bf{y}}} \right]{\rm{                                           }}
\end{cases}
\end{equation}
\end{footnotesize}where $S$ is a sparse matrix representing interference terms, which is updated alternatively with $\rho$ in the iterative process, ${\mathop{\rm mat}\nolimits} ( \cdot )$ is the inverse operator of ${\mathop{\rm mat}\nolimits} ( \cdot )$, ${D_\lambda }({\bf{X}})$ is the singular value contraction operator defined as ${D_\lambda }({\bf{X}}) = U{S_\lambda }(S){V^T}$, where $US{V^T}$ is the singular value decomposition of ${\bf{X}}$, and ${S_\lambda }({\bf{X}})$ is the soft threshold defined as \begin{footnotesize}${\left[ {{S_\lambda }({\bf{X}})} \right]_{ij}}\!=\! \begin{cases}
{x_{i{\kern 1pt} j}} - \lambda ,{\kern 1pt} {\rm{if}}{\kern 1pt} {\kern 1pt} {x_{i{\kern 1pt} j}} > \lambda \\
{x_{i{\kern 1pt} j}} + \lambda ,{\rm{if}}{\kern 1pt} {\kern 1pt} {x_{i{\kern 1pt} j}} < \lambda \\
0,{\kern 1pt} {\kern 1pt} {\kern 1pt} {\kern 1pt} {\kern 1pt} {\kern 1pt} {\kern 1pt} {\kern 1pt} {\kern 1pt} {\kern 1pt} {\kern 1pt} {\kern 1pt} {\kern 1pt} {\kern 1pt} {\kern 1pt} {\kern 1pt} {\kern 1pt} {\kern 1pt} {\kern 1pt} {\kern 1pt} {\kern 1pt} {\rm{otherwise}}
\end{cases}$\end{footnotesize}. $Y \in {R^m}$ is the Lagrange multiplier, and $\delta  \in [0, + \infty ]$ is the iterative step size, $\lambda ,\mu  > 0$. In the reconstruction experiments of this paper, the parameters of FP-ADMM algorithm are selected as follows: $\delta  = 1$, $\lambda  = 1/\sqrt d $ \cite{lab27}, $\mu  = 0.5/{\left\| {\bf{y}} \right\|_F}$, the initial values of $\rho$, $S$ and $Y$ are taken as zero matrices. The stopping criterion of the FP-ADMM algorithm is ${\left\| {{\kern 1pt} y - {\bf{A}} \cdot {\mathop{\rm vec}\nolimits} ({\rho ^k} + {{\bf{S}}^k})} \right\|_F}/{\left\| y \right\|_F} < {{\rm{\varepsilon }}_1}$ or the number of iterations $k > {k_{\max }}$,  let ${{\rm{\varepsilon }}_1} = {10^{ - 7}}$ and ${k_{\max }} = 30$.

In general, the process of reconstructing NMR quantum states with the method proposed can be summarized as follows: Randomly sample a certain number of $\left\{ {O_j^k} \right\}$ and $\left\{ {\left\langle {O_j^k} \right\rangle } \right\}$,  construct the convex optimization problem (2) with the sampled $\left\{ {O_j^k} \right\}$ and $\left\{ {\left\langle {O_j^k} \right\rangle } \right\}$, and solve (2) with the compressive FP-ADMM algorithm. The final optimal solution $\hat \rho $ is the reconstruction result of the state $\rho$.

The fidelity is used as the performance index of state reconstruction and is defined as:
\begin{equation}\label{eq7}
f = {{{\mathop{\rm Tr}\nolimits} \left( {\hat \rho {\rho ^\dag }} \right)} \mathord{\left/
 {\vphantom {{{\mathop{\rm Tr}\nolimits} \left( {\hat \rho {\rho ^\dag }} \right)} {\sqrt {{\mathop{\rm Tr}\nolimits} \left( {{{\hat \rho }^2}} \right){\mathop{\rm Tr}\nolimits} \left( {{\rho ^2}} \right)} }}} \right.
 \kern-\nulldelimiterspace} {\sqrt {{\mathop{\rm Tr}\nolimits} \left( {{{\hat \rho }^2}} \right){\mathop{\rm Tr}\nolimits} \left( {{\rho ^2}} \right)} }},
\end{equation}
where $\hat \rho $ and $\rho$ represent the experimentally reconstructed density matrix and the corresponding ideal density matrix, respectively, and $f \in [0,1]$.

\section{III. EXPERIMENTAL STATES RECONSTRUCTION IN NMR AND ANALYSIS}
We implement practical NMR experiments to reconstruct the states of $n = 2,3,4$ qubits, respectively, in order to examine the reduction performance of the number of measurements of our method. The experiments are carried out on a Bruker AV-400 spectrometer (9.4 T) at a room temperature of 303.0 K \cite{lab8}. The physical systems of $n = 2,3,4$ qubits states are $^13$C-labeled chloroform (${\rm{CHC}}{{\rm{L}}_{\rm{3}}}$) dissolved in deuterated acetone, Diethyl-fluoromalonate (${{\rm{C}}_7}{{\rm{H}}_{11}}{\rm{F}}{{\rm{O}}_4}$ ) dissolved in $^2$H-labeled chloroform, and iodotrifiuoroethylene (${{\rm{C}}_{\rm{2}}}{{\rm{F}}_{\rm{3}}}{\rm{I}}$) dissolved in d-chloroform, respectively. One $^1$H and one $^13$C are used for the first and second qubit of $n=2$, and one $^1$H, $^13$C and $^19$F are used for the first, second and third qubit of $n=3$. For $n=4$, one $^13$C is labeled as the first qubit, and $^19$F$_1$, $^19$F$_3$ and $^19$F$_3$ as the second, third, and fourth qubits, respectively. The systems are first prepared into pseudopure states (PPS) using the line selective-transition method \cite{lab32} in the experiment device. Then, by adjusting the pulse RF, the pseudo-pure states are manipulated into the target quantum state $\left| {{\psi _2}} \right\rangle $, $\left| {{\psi _3}} \right\rangle $ and $\left| {{\psi _4}} \right\rangle $ \citep{lab7,lab8}.

The state vectors associated with these three kinds of states are:
\begin{equation}\label{eq8}
\left| {{\psi _2}} \right\rangle  = \left| {00} \right\rangle,
\end{equation}
\begin{equation}\label{eq9}
\left| {{\psi _3}} \right\rangle  = \frac{4}{5}\left| {000} \right\rangle  - \frac{3}{5}\left| {001} \right\rangle,
\end{equation}
\begin{equation}\label{eq10}
\left| {{\psi _4}} \right\rangle  = \frac{1}{{\sqrt 2 }}\left( {\left| {0101} \right\rangle  + \left| {1010} \right\rangle } \right),
\end{equation}
in which $\left| 0 \right\rangle  = \left( {\begin{array}{*{20}{c}}
1\\
0
\end{array}} \right)$ and $\left| 1 \right\rangle  = \left( {\begin{array}{*{20}{c}}
0\\
1
\end{array}} \right)$ represent the ground state and the excited state of the nucleus, respectively. $\left| {{\psi _2}} \right\rangle $ is an eigenstate, and $\left| {{\psi _3}} \right\rangle $ and $\left| {{\psi _4}} \right\rangle $ are superposition states.

Let ${\rho _2} = \left| {{\psi _2}} \right\rangle \left\langle {{\psi _2}} \right|$, ${\rho _3} = \left| {{\psi _3}} \right\rangle \left\langle {{\psi _3}} \right|$ and ${\rho _4} = \left| {{\psi _4}} \right\rangle \left\langle {{\psi _4}} \right|$ be the corresponding density matrices of $\left| {{\psi _2}} \right\rangle $, $\left| {{\psi _3}} \right\rangle $ and $\left| {{\psi _4}} \right\rangle $. In order to accurately reconstruct ${\rho _2}$, ${\rho _3}$ and ${\rho _4}$, the states $\left| {{\psi _2}} \right\rangle $, $\left| {{\psi _3}} \right\rangle $ and $\left| {{\psi _4}} \right\rangle $ need to be prepared and observed repeatedly for the complete observation data. In practical NMR experiments, the total number of observable groups required are ${v_{_2}} = 6$, ${v_{_3}} = 16$ and ${v_{_4}} = 44$ for $n = 2,3,4$, respectively. The corresponding numbers of observables in each group are ${d_2} = {2^2} = 4$, ${d_3} = {2^3}= 8$ and ${d_4} = {2^4} = 16$. Thus, the total number of observables ${O_i}$ for $\left| {{\psi _2}} \right\rangle $, $\left| {{\psi _3}} \right\rangle $ and $\left| {{\psi _4}} \right\rangle $ are $6 \times 4 = 24$, $16 \times 8 = 128$ and $44 \times 16 = 704$, respectively, which are significantly larger than the theoretical number of complete measurement operators ${d^2}$, being 16, 64 and 256 for $n = 2,3,4$, respectively.

The ability of reconstructing quantum states using less sampling rate is significant for the proposed method. We do the experiments to demonstrate this ability in different cases. We carry out the experiments for 3 scenarios using two optimal algorithms and two kinds of sampling matrices for the comparisons:  Randomly sampling from the observable groups $\left\{ {O_j^k} \right\}$ by using (A) compressive FP-ADMM algorithm and (B) LS algorithm; (C) Randomly sampling from the measurement operators ${M_m}$ by using compressive FP-ADMM algorithm. The sampling rate ${\eta _g}$  in (3) is usually used to demonstrate the reduction performance of the number of the observable groups $\left\{ {O_j^k} \right\}$, and ${\eta _m} = {m \mathord{\left/
 {\vphantom {m {{d^2}}}} \right.
 \kern-\nulldelimiterspace} {{d^2}}}$ is used for the measurement operators $\left\{ {{M_m}} \right\}$. The performance of state reconstruction is the fidelity in (7). Under each sampling rate, we reconstruct each state 100 times and average over the resulting fidelities as the final average fidelity ${f_{avg}}$. The experimental results of reconstruction fidelities of ${\rho _2}$, ${\rho _3}$ and ${\rho _4}$ with different sampling rates in three cases are shown in Fig.1.

\begin{figure}[t]
\centering
\includegraphics[width=7.1cm]{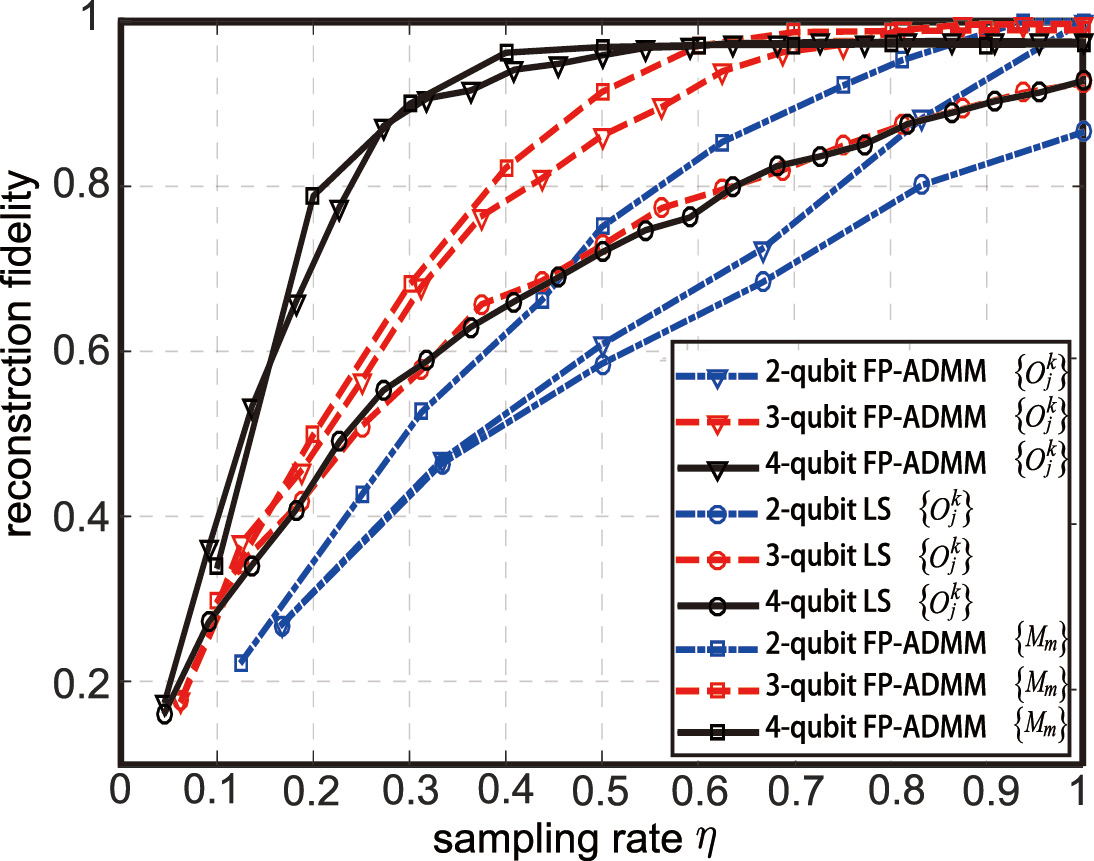}
\caption{The experimental results of reconstruction fidelities of  ${\rho _2}$, ${\rho _3}$ and ${\rho _4}$ with different sampling rates in three different cases. The blue dot-dash line, red dashed line and black solid line correspond to ${\rho _2}$, ${\rho _3}$ and ${\rho _4}$, and the triangle, circle and square mark correspond to the cases of A, B and C, respectively. For case A and B, the incremental step of sampling rates are selected as $\Delta {\eta _g} = 1/6,{\rm{ }}1/16{\rm{ }}$ and ${\rm{ }}1/22$ of ${\rho _2}$, ${\rho _3}$ and ${\rho _4}$, respectively, and for case C the incremental step of sampling rate is fixed as $\Delta {\eta _m} = 0.1$.}
\label{fig1}
\end{figure}

It can be seen from Fig. 1 that: All the reconstruction fidelities increases with the increase of the sampling rate. The average fidelities of FP-ADMM algorithm reach approximately 1 and remain stable when the corresponding sampling rates ${\eta _g}$  reach around 1, 0.75 and 0.5 with $n$ =2, 3 and 4, respectively. However, the maximum fidelities of LS with $n$ =2, 3 and 4 are only ${f_{avg - LS}} \sim 0.87,\; 0.93 \; and \; 0.93$ when ${\eta _m} = 1$. The compressive FP-ADMM algorithm is obviously better than the LS algorithm in the performance of state reconstruction.  For the two kinds of sampling matrices $\left\{ {O_j^k} \right\}$ and $\left\{ {{M_m}} \right\}$ using compressive FP-ADMM algorithm, the reconstruction fidelity of $\left\{ {O_j^k} \right\}$ is slightly worse than that of ${M_m}$ when $n=2$, but becomes close when $n=3$ and shows almost the same performance when $n=4$. This experimental result shows that the proposed method performs more efficiently with the increase of the system dimension size, which can be use to the state reconstruction of high-qubit quantum state in NMR.

The mean square error not only reflects the degree of discretization of the fidelities, but also responds to success probability of reconstruction at the corresponding sampling rate. We also do the experiments to study the mean square error of the fidelity at the different sampling rates of the proposed method with sampling matrices $\left\{ {O_j^k} \right\}$ using compressive FP-ADMM algorithm. Here we use $\zeta $ to represent the value of mean square error. In the experiments, we choose $f \ge 0.95$ as the criterion that the reconstruction is successful. When the average fidelity is near 0.95, the smaller of $\zeta $, the more concentrated the fidelity distribution, and the higher the success probability of reconstruction, and vice versa. The reconstruction average fidelity and mean square error $\zeta $ at different sampling rates using the proposed method are shown in Fig. 2.

\begin{figure}[t]
\centering
\includegraphics[width=7.1cm]{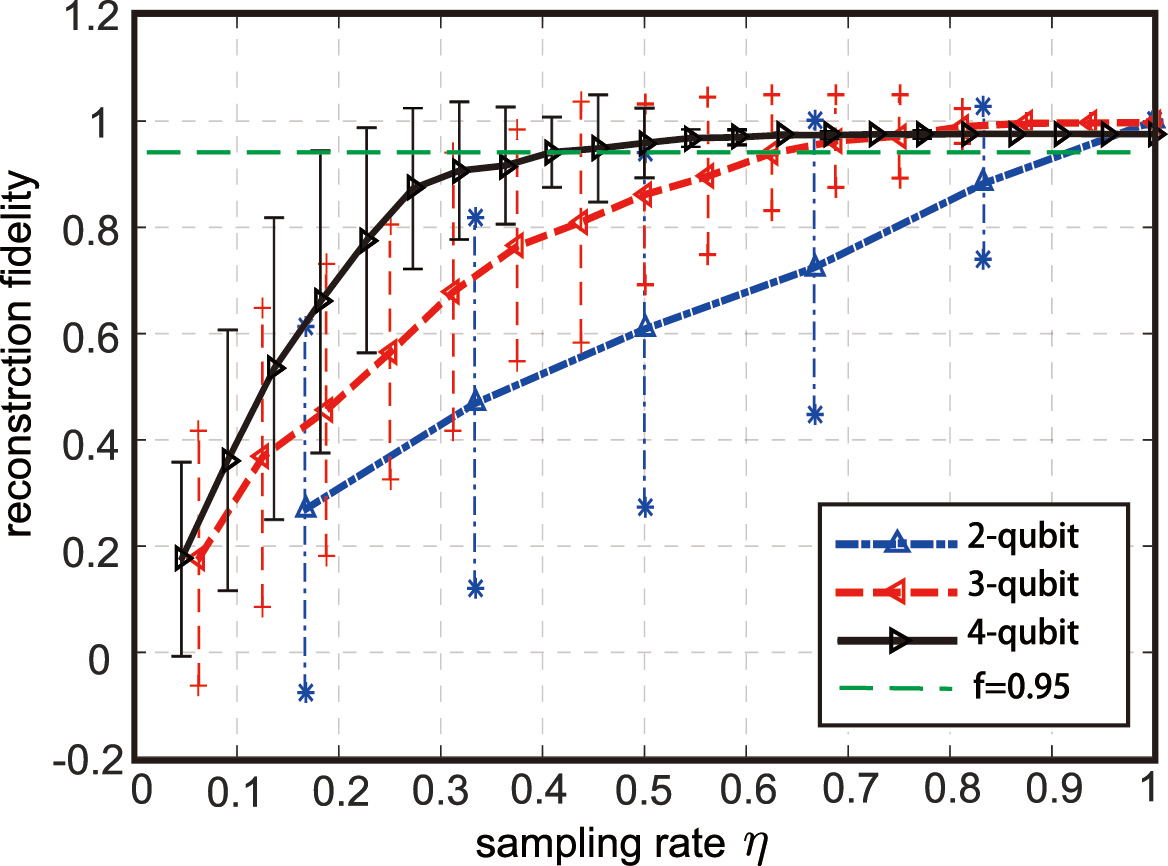}
\caption{The reconstruction average fidelity and mean square error $\zeta $ at different sampling rates using the proposed method. The blue dot-dash line with the inverted triangle, the red dash line with the left triangle, and the black solid line with the right triangle correspond to ${\rho _2}$, ${\rho _3}$ and ${\rho _4}$ respectively, while the star, plus and dash symbol represent the corresponding error bars. The green dash line represents the reconstruction fidelity $f=0.95$. The length of the error bar represents the mean square error of 100 experimental reconstruction fidelities. }\label{fig2}
\vspace{-1.2\baselineskip}
\end{figure}

Figure 2 shows that the mean square error $\zeta$ decreases with the increase of qubit number at the same sampling rate, e.g, the mean square errors are $\zeta  = 0.33,{\rm{ 0}}{\rm{.17 and 0}}{\rm{.07}}$ of n=2, 3 and 4 with ${\eta _g} = 0.5$.  $\zeta $  also tends to decrease with the increase of ${\eta _g}$ at the same qubit number.

We set the mean square errors $\zeta  \le 0.1$ to get a sufficiently large success probabilities of reconstruction (the probability that $f \ge 0.95$ ). The least sampling rates for $\zeta  \le 0.1$ of n=2, 3 and 4 are ${\eta _g} = 1,\;0.75\; and\;0.5$ , with the mean square errors being $\zeta  = 0,\;0.08 \; and \;0.07$, respectively. The least sampling rates are decreasing with the increase of qubit number. The average fidelities of reconstruction at these sampling rates are ${f_{avg}} = 1.0,\;0.97\;and\;0.96$  and the corresponding success probabilities of reconstruction are $100\% $, $92\%$ and $97\%$. This experimental results show that, we can carry out high-probability reconstruction of the quantum states in NMR with rather low sampling rates sing the proposed method, especially for high-qubit quantum states.

The experimental results of reconstructed density matrices of ${\rho _2}$, ${\rho _3}$ and ${\rho _4}$ are shown in Fig. 3, and the fidelities of the reconstructed density matrices in Figs. 3 (a) and (b) are shown in Table 1. In order to ensure a sufficiently high success probability of reconstruction, according to the experimental results of Fig. 2, we choose the sampling rates of ${\rho _2}$, ${\rho _3}$ and ${\rho _4}$ as ${\eta _{g2}} = 1.0$, ${\eta _{g3}} = 0.75$ and ${\eta _{g4}} = 0.50$, with the corresponding sampling rates being ${g_2} = 6$,  ${g_3} = 12$ and ${g_4} = 22$.

One can see from Table 1 that: The reconstruction fidelities by quantum state tomography are 0.9942, 0.9838 and 0.9606, respectively. And the reconstruction fidelities by the proposed method are 0.9999, 0.9896 and 0.9679, respectively, which have better performances than those of QST, indicating that our CS-QSR method is robust to the noise and interference in the actual measurement data to a certain extent. The important thing is the sampling rates (number of sampled groups) used in our method are also much less than 1 when the qubit number $n \ge 3$. The experimental results show that our method can reconstruct the actual NMR quantum states more accurately and effectively with only a small amount of observation data directly. The method proposed in this paper is the optimal reconstruction method under the existing conditions and can instruct the reconstructions of high-qubit quantum states in NMR.

\begin{figure}[t]
    \centering
    \begin{minipage}[t]{0.49\textwidth}
        \centerline{\includegraphics[width=\textwidth]{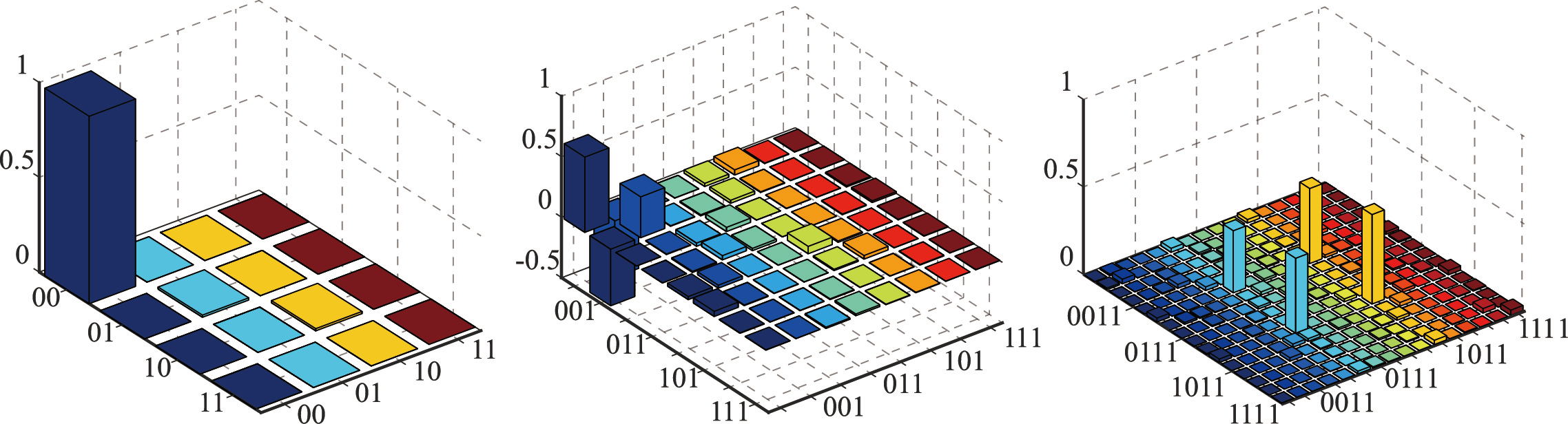}}
        \centerline{(a) reconstruction by means of QST with sampling rate ${\eta _m} = 1$ }\label{fig3-(a)}
    \end{minipage}
    \hfill
    \begin{minipage}[t]{0.49\textwidth}
        \centering{\includegraphics[width=\textwidth]{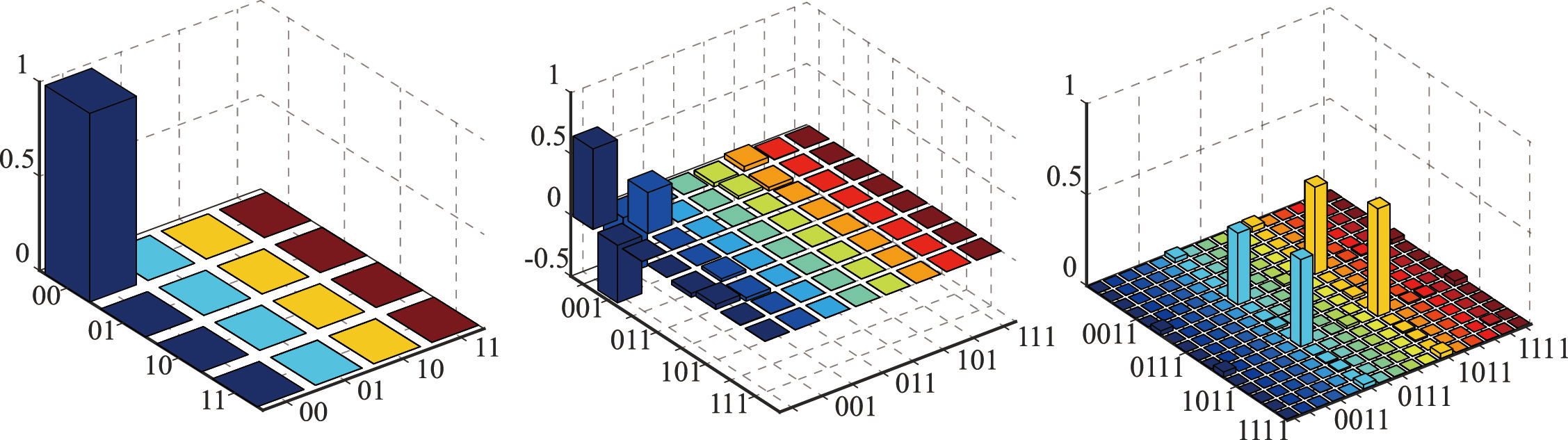}}
        \centering{(b) reconstruction by the proposed method with the sampling rates ${\eta _{g2}} = 1.0$, ${\eta _{g3}} = 0.75$ and ${\eta _{g4}} = 0.50$}\label{fig3-(b)}
    \end{minipage}
\caption{ The experimental results of reconstructed density matrices of ${\rho _2}$, ${\rho _3}$ and ${\rho _4}$. (a) is the reconstruction by means of QST with sampling rate ${\eta _m} = 1$, and (b) is the reconstruction by the proposed method. The three histograms from left to right in (a) and (b) correspond to the reconstructed density matrices of ${\rho _2}$, ${\rho _3}$ and ${\rho _4}$, respectively. Only the real parts of the reconstructed density matrices are given and the imaginary parts are ignored, because the imaginary parts of the elements in the ideal density matrices ${\rho _2}$, ${\rho _3}$ and ${\rho _4}$ are all 0.}\label{fig:3}
\end{figure}

\begin{table}[htb]
\centering
\begin{tabular}{|c|c|c|c|}
\hline
Fidelity & ${\rho _2}$ & ${\rho _3}$  & ${\rho _4}$ \\
\hline
QST  & 0.9942 & 0.9838 & 0.9606   \\
\hline
Our method & 0.9999 & 0.9896 & 0.9679 \\
\hline
\end{tabular}
\caption{The fidelities of the reconstructed density matrices in Fig. 3 (a) and (b). }\label{tab1}
\end{table}

\section{IV. CONCLUSION}
In this paper, we first reconstructed actual NMR quantum states via compressive sensing. We also proposed an effective NMR quantum state reconstruction method based on CS and gave a detailed derivation of the method in both theoretical and experimental aspects. The observation data is directly used in our method so as to save the transformation process of QST, which effectively enhances the efficiency of the state reconstruction in practical NMR experiments. We validated our method with actual observation data of different qubit states and analyzed the effect of different factors on the reconstruction performance. The method proposed in this paper is both feasible in implementation and accurate in reconstruction and can greatly reduce the number of measurements required, which provides a new protocol for the state reconstruction with higher qubits in practical NMR experiments.

\section{ACKNOWLEDGMENTS}

This work was supported by the National Natural Science Foundation of China (61573330).

\bibliographystyle{apsrev4-1} 

\end{document}